\documentclass[a4paper,twocolumn,showpacs,superscriptaddress,floatfix]{revtex4}
\usepackage{graphicx}
\usepackage{epsf}
\usepackage{epsfig}
\usepackage{amssymb,amsmath}
\usepackage[usenames]{color}
\usepackage{amssymb}
\usepackage{times}

\newcommand{\vk}{v_{\rm k}}
\newcommand{\ks}{\,{\rm km/s}}
\newcommand{\cf}{\textit{cf.}~}
\newcommand{\ie}{\textit{i.e.,}~}

\begin{document}

\title{Understanding the ``anti-kick'' in the merger of binary black
  holes}

\author{Luciano Rezzolla}
\affiliation{
Max-Planck-Institut f{\"u}r Gravitationsphysik, Albert Einstein
Institut, Potsdam, Germany 
}
\affiliation{
Department of Physics and Astronomy,
Louisiana State University,
Baton~Rouge, Louisiana, USA
}

\author{Rodrigo P. Macedo}
\affiliation{
Max-Planck-Institut f{\"u}r Gravitationsphysik, Albert Einstein
Institut, Potsdam, Germany 
}
\affiliation{
Instituto de F\'\i sica,  Universidade de S\~ao Paulo, 
S\~ao Paulo, SP, Brazil
}

\author{Jos\'e Luis Jaramillo}
\affiliation{
Max-Planck-Institut f{\"u}r Gravitationsphysik, Albert Einstein
Institut, Potsdam, Germany 
}
\affiliation{
Laboratoire Univers et Th\'eories, Observatoire de Paris, CNRS, Meudon, France
}

\begin{abstract}
The generation of a large recoil velocity from the inspiral and merger
of binary black holes represents one of the most exciting results of
numerical-relativity calculations. While many aspects of this process
have been investigated and explained, the ``antikick'', namely the
sudden deceleration after the merger, has not yet found a simple
explanation. We show that the antikick can be understood in terms of
the radiation from a deformed black hole where the 
anisotropic curvature distribution on the horizon correlates with the
direction and intensity of the recoil. Our analysis is focussed on
Robinson-Trautman spacetimes and allows us to measure both the
energies and momenta radiated in a gauge-invariant manner. At the same
time, this simpler setup provides the qualitative and quantitative
features of merging black holes, opening the way to a deeper
understanding of the nonlinear dynamics of black-hole spacetimes.


\end{abstract}

\pacs{04.30.Db, 04.25.dg, 04.70.Bw, 97.60.Lf}

\maketitle

\noindent\emph{Introduction.~}The merger of two black holes (BHs) is
one of the most important sources of gravitational waves (GW) and it
is generally accompanied by the recoil of the final BH as a result of
anisotropic GW emission. While this scenario has been investigated for
decades~\cite{peres:1962} and first estimates have been made using
approximated and semianalytical methods such as a particle
approximation~\cite{1984MNRAS.211..933F}, post-Newtonian
methods~\cite{Wiseman:1992dv} and the close-limit approximation
(CLA)~\cite{Andrade:1997pc}, it is only recently that accurate values
for the recoil have been computed~\cite{Baker:2006nr, Gonzalez:2006md,
  Campanelli:2007cg, Herrmann:2007ac, Koppitz-etal-2007aa:shortal,
  Lousto:2007db, Pollney:2007ss:shortal, Healy:2008js}.

Besides being a genuine nonlinear effect of general relativity, the
generation of a large recoil velocity during the merger of two BHs has
a direct impact in astrophysics. Depending on its size and its
variation with the mass ratio and spin, in fact, it can play an
important role in the growth of supermassive BHs via mergers of
galaxies and on the number of galaxies containing
BHs~\cite{merritt08}. Numerical-relativity simulations of BHs
inspiralling on quasicircular orbits have already revealed many of
the most important features of this process showing, for instance,
that asymmetries in the mass can lead to recoil velocities $\vk
\lesssim 175\ks$~\cite{Baker:2006nr, Gonzalez:2006md}, while
asymmetries in the spins can lead respectively to $\vk \lesssim
450\ks$ or $\vk \lesssim 4000\ks$ if the spins are
aligned~\cite{Herrmann:2007ac,Koppitz-etal-2007aa:shortal,
  Pollney:2007ss:shortal} or perpendicular to the orbital angular
momentum~\cite{Campanelli:2007ew, Gonzalez:2007hi,Campanelli:2007cg}
(see~\cite{Rezzolla:2008sd} for a review).

At the same time, however, there are a number of aspects of the
nonlinear processes leading to the recoil that are far from being
clarified even though interesting work has been recently carried out
to investigate such aspects~\cite{Schnittman:2007ij, Mino:2008at,
  LeTiec:2009yg}. One of these features, and possibly the most
puzzling one, is the generic presence of an ``antikick'', namely, of
one (or more) decelerations experienced by the recoiling BH. Such
antikicks take place after a single apparent horizon (AH) has been
found and have been reported in essentially all of the mergers
simulated so far.

This Letter is dedicated to elucidate the stages during which the
antikick is generated and to provide a simple and qualitative
interpretation of the physics underlying this process. Our focus will
be on the head-on collision of two nonspinning BHs with different mass
and although this is the simplest scenario for a BH-merger, it
contains many of the aspects that can be encountered in more generic
conditions. Our qualitative picture will then be made quantitative and
gauge-invariant by studying the logical equivalent of this process in
the evolution of a Robinson-Trautman (RT) spacetime, with measurements
of the recoil made at future null infinity. As commented below, the
insight gained with RT spacetimes will be valuable to explain the
antikick under generic conditions.

\begin{figure}[t]
\begin{center}
\includegraphics[width=0.3\textwidth,angle=-90]{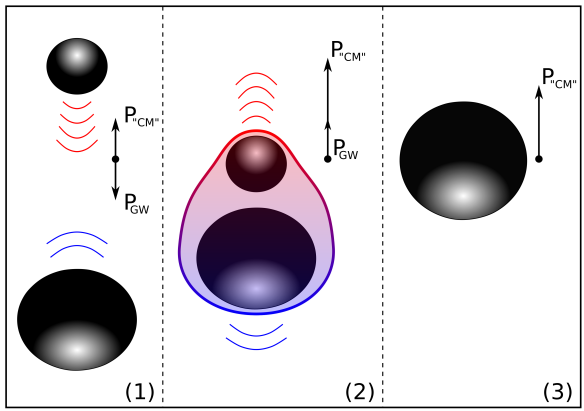}
\end{center}
\vglue-.25cm
\caption{Cartoon of the generation of the antikick in the head-on
  collision of two unequal-mass Schwarzschild BHs. Initially the
  smaller BH moves faster and linear momentum is radiated mostly
  downwards, thus leading to an upwards recoil of the system [stage
    (1)]. At the merger the curvature is higher in the upper
  hemisphere of the distorted BH (\cf shading) and linear momentum is
  radiated mostly upwards leading to the antikick [stage (2)]. The BH
  decelerates till a uniform curvature is restored on the horizon
  [stage (3)].}  \vglue-0.5cm
\label{fig:fig0}
\end{figure}

\smallskip\noindent\emph{The basic picture.~}Before discussing how to
use the RT spacetime to compute the antikick, it is useful to
illustrate the basic BH physics leading to such process and for this
we consider the collinear merger of two Schwarzschild BHs with unequal
masses. This is shown in a schematic cartoon in Fig.~\ref{fig:fig0},
where we have considered a reference frame in the centre of mass of
the system and where the smaller black hole is initially on the
positive $z$ axis, while the larger one is on the negative axis. As
the two BHs free-fall towards each other, the smaller one will move
faster and will be more efficient in ``forward-beaming'' its GW
emission~\cite{Wiseman:1992dv}. As a result, the linear momentum will
be radiated mostly downwards, thus leading to an upwards recoil of the
BH binary [\cf stage (1) in Fig.~\ref{fig:fig0}]. At the merger the BH
velocities will be larger and so will also be the anisotropic GW
emission and the corresponding recoil of the system. However, when a
single AH is formed comprising the two BHs, the curvature distribution
on this 2-surface will be highly anisotropic, being higher in the
upper hemisphere [\cf shading in stage (2) of
  Fig.~\ref{fig:fig0}]. Because the newly formed BH will want to
radiate all of its deviations away from the final Schwarzschild
configuration, it will do so more effectively there where the
curvature is larger, thus with a stronger emission of GWs from the
northern hemisphere. As a result, after the merger the linear momentum
will be emitted mostly upwards and this sudden sign change will lead
to the antikick. The anisotropic GW emission will decay exponentially
as the curvature gradients are erased and the BH will have reached its
final and decelerated recoil velocity [\cf stage (3)].

Although this picture refers to a head-on collision, it is supported
by the findings in the CLA (where the direction of the ringdown kick
is approximately opposite to that of the accumulated inspiral plus
plunge kick)~\cite{LeTiec:2009yg} and it can be generalized to a
situation in which the BHs have different masses, different spins and
are merging through an inspiral. Also in a more generic case, in fact,
the newly formed AH will have a complicated but globally anisotropic
distribution of the curvature, determining the direction (which is in
general varying in time) along which the GWs will be emitted.
Therefore we argue that the geometric properties in a dynamical
horizon (of a black or white hole) determine its global dynamics.  We
next use the RT spacetime to validate this picture.

\smallskip\noindent\emph{The Robinson-Trautman spacetime.~} It is a
class of vacuum solutions admitting a congruence of null geodesics
which are twist and shear-free~\cite{Robinson:1962zz}, with a future
stationary horizon and a dynamical past (outer trapping)
horizon~\cite{PenroseTod} (past AH hereafter). A RT spacetime can thus
be regarded as an isolated nonspherical white hole emitting GWs, where
the evolution of the AH curvature-anisotropies and of the spacetime
momentum can be related unambiguously. The metric
is~\cite{Macedo:2008ia}
\begin{equation}
\label{RTM}
ds^2 = -\left(K - \frac{2M_{\infty}}{r} -
\frac{2r \partial_u Q}{Q} \right) du^2 - 2dudr + \frac{r^2}{Q^2} d\Omega^2,
\end{equation}
where $Q=Q(u,\Omega)$, $u$ is the standard null coordinate, $r$ is the
affine parameter of the outgoing null geodesics, and $\Omega=\{
\theta, \phi \}$ are the angular coordinates on the unit sphere
$S^2$. Here $M_{\infty}$ is a constant and is related to the
asymptotic mass, while the function $K(u,\Omega)$ is the Gaussian
curvature of the surface corresponding to $r = 1$ and $u = {\rm
  constant}$, $K(u,\Omega)\equiv Q^2(1+\nabla^2_{\Omega}{\ln{Q}})$,
where $\nabla^2_{\Omega}$ is the Laplacian on $S^2$. The Einstein
equations then lead to 
\begin{equation}
\label{RTeq}
\partial_u Q(u,\Omega)=-{Q^3} \nabla^2_{\Omega}K(u,\Omega)/({12M_{\infty}}).
\end{equation}
Any regular initial data $Q=Q(0,\Omega)$ will smoothly evolve
according to~\eqref{RTeq} until it achieves a stationary configuration
corresponding to a Schwarzschild BH at rest or moving with a constant
speed~\cite{Chrusciel:1992cj}.  Equation~\eqref{RTeq} implies the
existence of the constant of motion $\mathcal{A}\equiv
\int_{S^2}{d\Omega}/{Q^2}$, which clearly represents the area of the
surface $u,r=$ constant and can be used to normalise $Q$ so that
$\mathcal{A}=4\pi$.  All the physically relevant information is
contained in the function $Q(u,\Omega)$, and this includes the
gravitational radiation, which can be extracted by relating
$Q(u,\Omega)$ to the radiative part of the Riemann
tensor~\cite{deOliveira:2004bn,Aranha:2008ni}.

The past AH radius $R(u,\Omega)$ is given by the vanishing expansion
of the future ingoing null geodesics~\cite{PenroseTod}
\begin{equation}
\label{BHRT}
Q^2\nabla_{\Omega}^{2} \ln{R}=K-{2M_{\infty}}/{R}.
\end{equation}

The mass and momentum of the BH are computed at future null infinity
using the Bondi 4-momentum~\cite{Macedo:2008ia}
\begin{equation}
\label{BondiMoment}
P^{\alpha}(u) \equiv
\frac{M_{\infty}}{4\pi}\int_{S^2}\frac{\eta^{\alpha}}{Q^3}d\Omega,
\end{equation}
with $\left\{\eta^{\alpha}\right\}= \left\{1, \sin{\theta}\cos{\phi},
\sin{\theta}\sin{\phi}, \cos{\theta} \right\}$. Given smooth initial
data, the spacetime will evolve to a stationary nonradiative solution
which, in axisymmetry, has the form $Q(\infty, \theta)= {\left(1 \mp v
  x\right)}/{\sqrt{1-v^2}}$, with $x\equiv
\cos\theta$~\cite{Macedo:2008ia}. The Bondi 4-momentum associated to
$Q(\infty, \theta)$ is
\begin{equation}
\label{Bondiinfty}
\left\{P(\infty)\right\}^{\alpha}= \left({M_{\infty}}/{\sqrt{1-v^2}}\right)
\left\{1,0,0,\pm v \right\},
\end{equation}
so that the parameter $v$ in $Q(\infty, \theta)$ can be interpreted as
the velocity of the Schwarzschild BH in the $z$-direction.

One of the difficulties with RT spacetimes is the definition of
physically meaningful initial data. Although we are more interested in
a proof of principle than in a realistic configuration, we have
adopted the prescription in~\cite{Aranha:2008ni}
\begin{equation}
\label{Q0HeadOn}
Q(0,\theta)=Q_0\left[\frac{1}{\sqrt{1 - w x}}+
\frac{q}{\sqrt{1 + w x}} \right]^{-2},
\end{equation}
which was interpreted to represent the final stages (\ie after a
common AH is formed) of a head-on collision of two boosted BHs with
opposite velocities $w$ and mass ratio $q$~\cite{Aranha:2008ni}. In
practice, to reproduce the situation shown in Fig.~\ref{fig:fig0}, we
have set $w<0$ and taken $q\in [0,1]$, but a more general class of
initial data can be easily constructed.  Note that $Q_0$ is normalized
so that to $\mathcal{A}=4\pi$ and that in general the deformed BH will
not be initially at rest. As a result, given the initial velocity
$v_0\equiv P^3(0)/P^0(0)$, we perform a boost $\overline{P}^{\alpha} =
\Lambda^{\alpha}_{~\beta}(v_0)P^{\beta}$ so that
$\overline{P}^{3}(0)=0$ by construction. The numerical solution of
Eq.~\eqref{RTeq} with initial data~\eqref{Q0HeadOn} is performed as
discussed in~\cite{Macedo:2008ia}.

\begin{figure}[t]
\begin{center}
\includegraphics[width=8.0cm,clip=true]{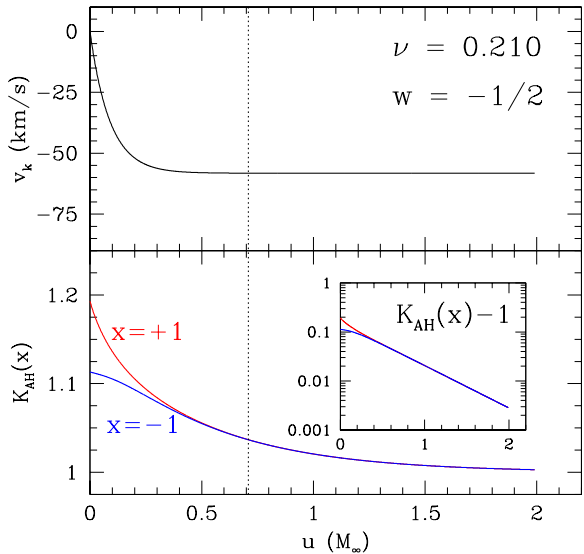}
\end{center}
\vglue-0.75cm
\caption{Typical evolution of a RT spacetime. Shown in the lower panel
  is the evolution of the curvature $K_{_{\rm AH}}$ at the north
  ($x=1)$ and south pole ($x=-1)$. Shown in the upper panel is the
  evolution of the recoil, which stops decreasing when the curvature
  difference is erased by the radiation (dotted line).}  \vglue-0.5cm
\label{fig:fig3}
\end{figure}

\smallskip\noindent\emph{Discussion.~}Figure~\ref{fig:fig3} reports
the typical evolution of a RT spacetime with the lower panel showing
the evolution of the curvature of the past AH $K_{_{\rm AH}}\equiv
2M_{\infty}/R^3(x)$ at the north ($x = 1)$ and south pole ($x=-1)$,
and with the upper panel showing the evolution of the recoil
velocity. Note that the two local curvatures are different initially,
with the one in the upper hemisphere being larger than the one in the
lower hemisphere (\cf~Figure~\ref{fig:fig0}). However, as the
gravitational radiation is emitted, this difference is erased. When
this happens, the deceleration stops and the BH attains its asymptotic
recoil velocity. The inset reports the curvature difference relative
to the asymptotic Schwarzschild one, $K_{_{\rm AH}}-1$, whose
exponentially decaying behaviour is the one expected in a ringing BH.

\begin{figure}[t]
\begin{center}
\includegraphics[width=8.0cm,clip=true]{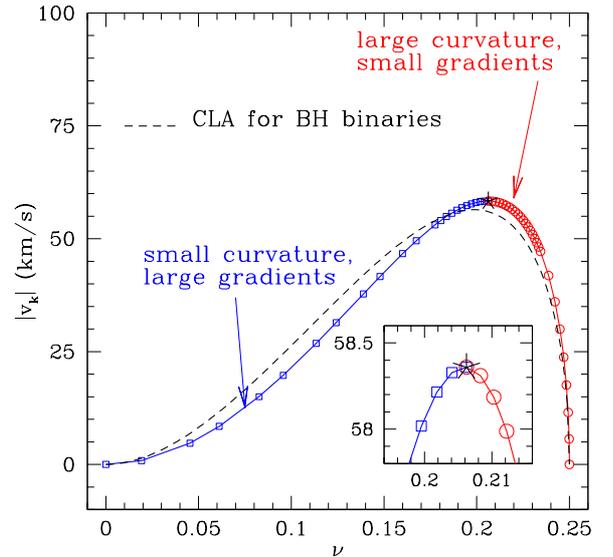}
\end{center}
\vglue-0.75cm
\caption{Recoil velocity shown as a function of the symmetric mass
  ratio $\nu$ when $w=-0.425$, with the dashed line refers to the
  antikick from the inspiral of nonspinning binaries in the
  CLA~\cite{LeTiec:2009yg}. Note that the curve can be thought of as
  being composed of two different branches.}
\label{fig:fig1}
\vglue-0.5cm
\end{figure}

As mentioned before, that shown in Fig.~\ref{fig:fig3} is a typical
evolution of a RT spacetime and is not specific of the initial
data~\eqref{Q0HeadOn}. By varying the values of $w$, in fact, it is
possible to increase or decrease the final recoil, while a sign change
in $w$ simply inverts the curvature at the poles so that, for
instance, initial data with $w>0$ would yield a BH accelerating in the
positive $z$-direction. Interestingly, it is even possible to
fine-tune the parameter $w$ so that the recoil produced for a RT
spacetime mimics the antikick produced by the quasicircular inspiral
of nonspinning binaries. This is shown in Fig.~\ref{fig:fig1}, which
reports the recoil as a function of the symmetric mass ratio
$\nu\equiv q/(1+q)^2$, and where the dashed line refers to the
antikick for the inspiral of nonspinning binaries in the
CLA~\cite{LeTiec:2009yg} (the parameters chosen, \ie $w=-0.425$ and
$r_{12}=2\,M$, are those minimizing the differences). Considering that
the two curves are related only logically and that the CLA one
contains all the information about inspiralling BHs, including the
orbital rotation, the match is surprisingly good.

It is also suggestive to think that the curve in Fig.~\ref{fig:fig1}
is actually composed of two different branches, one of which is
characterized by large curvature gradients across the AH but small
values of the curvature (this is the low-$\nu$ branch and is indicated
with squares), while the other is characterized by small curvature
gradients and large values of the curvature (this is the high-$\nu$
branch and is indicated with circles). The same recoil velocity can
then be produced by two different values of $\nu$, for which the
effects of large curvature gradients and small curvatures are the same
as those produced by small curvature gradients but large curvatures.

To go from this intuition to a mathematically well-defined measure we
have computed the mass multipoles of the intrinsic curvature of the
initial data using the formalism developed in~\cite{Ashtekar04a} for
dynamical horizons. Namely, we have calculated the mass moments as
\begin{equation}
\label{curv_multipoles}
M_n \equiv \oint \frac{P_{n}(\tilde{x})}{Q^2(\theta)R(\theta)}d\Omega,
\end{equation}
where $P_{n}(\tilde{x})$ is the Legendre polynomial in terms of the
coordinate $\tilde{x}(\theta)$ which obeys
$\partial_{\theta}\tilde{x}=- \sin\theta R(\theta)^2/(R_{_{\rm AH}}^2
Q(\theta)^2)$, with $R_{_{\rm AH}}\equiv \sqrt{{\cal A}_{_{\rm
      AH}}/(4\pi)}$ and $\tilde{x}(0)=1$.  Using these multipoles it
is possible to construct an effective-curvature parameter $K_{\rm
  eff}$ that represents a measure of the global curvature properties
of the initial data and from which the recoil depends in an injective
way. Because this effective-curvature parameter has to contain the
contribution from the even and odd multipoles, we have found that the
expression $K_{\rm eff} = M_2|\sum_{n=1} M_{2n+1}/3^{n-1}|$,
reproduces exactly what is expected (note $M_1=0$ to machine precision).

\begin{figure}[t]
\begin{center}
\includegraphics[width=8.0cm,clip=true]{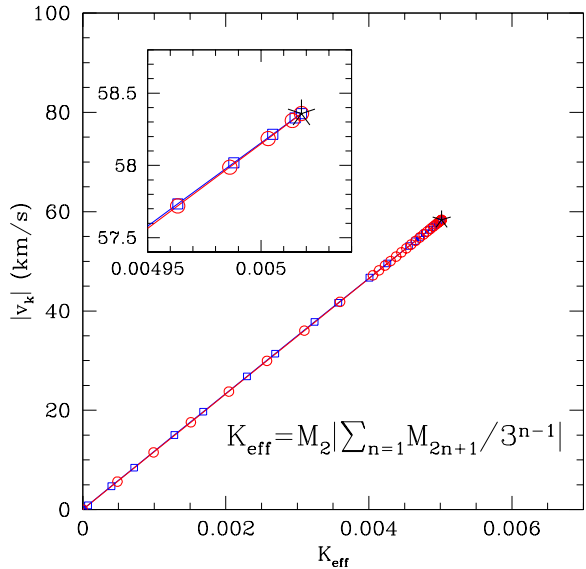}
\end{center}
\vglue-0.75cm
\caption{Recoil velocity shown as a function of the effective
  curvature. In contrast with Fig.~\ref{fig:fig1}, which uses the same
  symbols employed here, the relation between the curvature and the
  recoil is now injective.}  \vglue-0.5cm
\label{fig:fig4}
\end{figure}

This is shown in Fig.~\ref{fig:fig4}, which reports the recoil
velocity as a function of $K_{\rm eff}$. As predicted, and in contrast
with Fig.~\ref{fig:fig1}, the relation between the curvature and the
recoil is now injective, with the maximum recoil velocity being given
by the maximum value of $K_{\rm eff}$ (see inset), and with the two
branches coinciding. We do not expect the expression found here for
$K_{\rm eff}$ to be unique and indeed a more generic one will have to
include also the mass-current multipoles to account for the spin
contributions. However, lacking a rigorous mathematical guidance, our
phenomenological $K_{\rm eff}$ is a reasonable, intuitive
approximation.

\smallskip\noindent\emph{Conclusions.~}We have outlined a simple
picture to explain the deceleration observed during the merger of
binary BHs in terms of the dissipation of an anisotropic distribution
of curvature on the horizon of the newly formed BH. We have analyzed
this picture for the head-on collision of two nonspinning BHs with
unequal mass but its extension to generic systems is direct as the
same features will be present also when including the spin and the
orbital contributions: mass-current multipoles will add (substract) in
prograde (retrograde) orbits. The qualitative arguments made on the
head-on collision have then been made quantitative by analyzing the
gauge-independent dynamics of RT spacetimes. More specifically we have
shown that the deceleration is associated to the radiation of
curvature differences and persists as long as the gradients are not
erased. Furthermore, the directionality of the recoil is dictated by
the north-south curvature gradients and a one-to-one mapping between
the recoil and an effective curvature is possible. These results
presented here can help in understanding some nonlinear aspects of 
curved spacetimes.

Finally, an alternative interpretation of the recoil phenomenology can
be given via the Landau-Lifshitz pseudotensor, where the recoil is
given by the cancellation of large and opposite fluxes of momentum,
part of which are ``swallowed'' by the
BH~\cite{Lovelace:2009dg}. While this is an interesting route, it
relies on gauge-dependent measurements which may themselves be
counterintuitive.


\smallskip\noindent\emph{Acknowledgments.~} It is a pleasure to thank
A. Saa, B. Krishnan, H. Oliveira, B. Schutz, and I. Soares for useful
discussions. We are also grateful to A. Le Tiec for providing his
estimates of the recoil in the CLA. This work was supported in part by
the DAAD and the DFG grant SFB/Transregio~7.



\end{document}